\documentclass[]{aa}  
\usepackage{graphicx}
\usepackage{color}
\usepackage{subcaption}         
\usepackage{lscape}             
\usepackage{placeins}           
\definecolor{kh}{cmyk}{1.0, 0.8, 0., 0.15}
\def\kh#1{\textcolor{kh}{#1}}

\usepackage{txfonts}
\usepackage{amsmath}
\usepackage{hyperref}

\begin{document} 

\title{Comparing the Sun to Sun-like stars}
\subtitle{On the importance of addressing faculae/spot domination}

   \author{Konstantin Herbst
          \inst{1,2}
          \and
          Eliana M. Amazo-G\'{o}mez\inst{3}
          \and
          Athanasios Papaioannou\inst{4}
          }

   \institute{Institut f\"ur Planetenforschung (PF). Deutsches Zentrum f\"ur Luft- und Raumfahrt (DLR). Rutherfordstr. 2. 12489 Berlin. Germany
        \and
            Centre for Planetary Habitability (PHAB). Department of Geosciences. University of Oslo. Sem Sæland 2A. 0371 Oslo. Norway\\
              \email{konstantin.herbst@geo.uio.no}
        \and
            Leibniz-Institut f\"ur Astrophysik Potsdam. An der Sternwarte 16. 14482 Potsdam. Germany
        \and
             Institute for Astronomy. Astrophysics. Space Applications and Remote Sensing (IAASARS). National Observatory of Athens. I. Metaxa \& Vas. Pavlou St. 15236 Penteli. Greece
             }

   \date{}
 
  \abstract
   {Whether the Sun is an ordinary G-type star is still an open scientific question. Stellar surveys by Kepler and TESS, however, revealed that Sun-like stars tend to show much stronger flare activity than the Sun.}
   {This study aims to reassess observed flare and spot activity of Sun-like Kepler stars by fine-tuning the criteria for a more robust definition of Sun-like conditions and better comparability between the current Sun and Sun-like stars.}
   {We update one of the recent stellar Sun-like star samples by applying new empirical stellar relations between the starspot size and the effective stellar temperature to derive more reliable starspot group sizes. From the 265 solar-type stars, we could select 48 stars supporting the Kepler 30-minute cadence light curves. These were analyzed by implementing the gradient of the power spectra method to distinguish between spot- and faculae-dominated stars. We employed the $\alpha$-factor to quantify the area ratio of bright and dark features on the stellar surface.}
   {We were able to group the 48 stars as being spot- or faculae-dominated, revealing a preferential distribution of the Kepler Sun-like stars towards the spot-dominated (44 stars) and transitional (four stars) regimes. As the current Sun is faculae-dominated, only the transitional stars were utilized for further evaluation. Additionally, accounting for comparability in stellar mass, radius, and rotation period, we show that only one of the utilized 265 Sun-like stars in the Kepler sample (i.e., KIC\,11599385) allows for direct comparison to the current Sun. We further show that the single flare observed on KIC\,11599385 falls right within the flare energy range estimated for the AD774/775 event observed in the cosmogenic radionuclide archives of $^{10}$Be, $^{14}$C, and $^{36}$Cl. 
   }
   {}


   \maketitle
%

\section{Introduction} \label{sec:intro}
With an age of approximately 4.6 Grs and a rotation rate of about 25 days, our Sun is an old and slowly rotating main sequence G-type star. However, it remains magnetically active, often producing eruptive events such as flares and coronal mass ejections. These events vary by several orders of magnitude in energy and occur across a wide range of timescales. The maximum threshold of such events is not well constrained and, until recently, has been limited to direct instrumental observations of the past $\sim$ 50 years. In addition, strong solar energetic particle events, such as ground-level enhancements (GLEs), have been observed, occurring on average once per year \citep{macau_mods_00003837}. With the help of cosmogenic radionuclide records (i.e., of $^{10}$Be, $^{14}$C, and $^{36}$Cl), also much more extreme solar energetic particle events (ESPE) were identified on a multi-millennial timescale \citep[e.g.,][]{2012Natur.486..240M,mekhaldi2015multi}.

Missions such as Kepler \citep[e.g.,][]{2010ApJ...713L..79K} and TESS \citep[e.g.,][]{RickerEA2015} observed flares on other stars. These stellar flares \citep[e.g.,][]{2005stam.book.....G, 2005nlds.book.....R, 2022LRSP...19....2C} are believed to result from sudden discharges of magnetic energy accumulated around starspots, similar to what we know from solar flares \citep[e.g.,][]{2011LRSP....8....6S}. Thereby, the observations showed that (i) young stars with rotation rates of only a few days, (ii) binary stars, and (iii) cool M stars are even more magnetically active and tend to produce so-called superflares \citep[e.g.,][]{BenzGüdel2010, TristanEA2023} that exceed solar bolometric flare energies by a factor of 10 to 10$^6$ \citep[i.e., $\sim 10^{33} - 10^{38}$ erg][]{Schaefer_2000}.

Using data from the space-borne Kepler mission, \citet{2012Natur.485..478M} analyzed a sample of 83,000 solar-type stars — specifically G-type main-sequence stars with effective temperatures (T$_{\mathrm{eff}}$) between 5,100 K and 6,000 K and surface gravity log(g) values of at least 4.0. They identified 365 superflares occurring on 148 solar-type stars, showing bolometric energies at least an order of magnitude higher than typical solar flares \citep[$\sim 10^{32}$erg,][]{Emslie2012}.

By investigating more extended Kepler periods in the order of 500 days, later \citet{2013ApJS..209....5S} found 1547 flares on 279 solar-type stars, revealing that the bolometric flare energies of solar-type stars are 10 to 10$^4$ orders of magnitude stronger (i.e., $10^{33}$ to $10^{36}$ erg). But the release of the Gaia-DR2 stellar radius data in 2018 indicated potential contamination of the sub-giants in the Kepler sample of solar-type stars. Accordingly, \citet{Notsu-etal-2019} re-investigated the \citet{2013ApJS..209....5S} sample and found 40\% of the solar-type stars to be sub-giants that thus needed to be removed from the statistical sample. This drastically reduced the number of stars in the stellar solar-type sample, and with that the sample of Sun-like stars, a subset of solar-type stars with effective temperatures ranging between 5,600 K and 6,000 K, $\log(g)$ $\ge 4.0$, and rotation periods $>$ 20 days \citep[i.e., slow rotators, see, e.g., Tab. 1 in][]{OkamotoEA2021}.

Taking into account all data of the four-year Kepler mission, covering about 1500 days of observations, the Gaia-DR2 results, and updating the method to detect flares in the sample recently \citet{OkamotoEA2021} derived a new subsequent statistical analysis. The dedicated work finally led to an increase in the number of solar-type stars and particular Sun-like stars compared to the sample by \citet{Notsu-etal-2019}. This new sample allowed a more dedicated statistical analysis, revealing 2341 superflares on 265 solar-type stars and 26 on 15 Sun-like stars. In addition, \citet{OkamotoEA2021} showed that (i) flare energies of up to 10$^{36}$ erg occurred on young (a few hundred Myr) and fast-rotating (rotation period $P_{\mathrm{rot}}$ around a few days) solar-type stars, (ii) the flare energy decreases with increasing rotation period in the case of solar-type stars, and (iii) superflares with energies of up to $4\cdot 10^{34}$ erg occurred on Sun-like stars. Statistically speaking, Sun-like stars show superflares with energies of around 7$\cdot 10^{33}$ erg (10$^{34}$ erg) on average every 3000 (6000) years \citep[e.g.,][]{Hayakawa2023,UsoskinEA2023}. Note that their methods are bound to stars with known stellar rotation periods.

However, when it comes to flaring stars, a crucial stellar question  -- so far -- has not been addressed in all of these statistical studies: Are these Sun-like stars spot-dominated or faculae-dominated like the current Sun (and therewith directly comparable with it)?

Bright magnetic features such as faculae, plages, and networks, as well as dark magnetic features like spots, have been extensively differentiated and analyzed in the Sun. These differences are attributed to the relative thickness of the emerging magnetic flux tubes \cite[e.g.,][]{2000JApA...21..275F,2002JASTP..64..677S}, with specific references to the dimensions of the structure. However, in the context of other stars, obtaining comparable data in both quantity and quality is currently unrealistic. The lack of continuous observations and unresolved surface details makes distinguishing faculae from spots particularly challenging. 

The center-to-limb contrast modulates the brightness distribution of stellar disks, manifesting as the limb-darkening effect in the continuum. This optical phenomenon results in the central region appearing brighter than the edges of the disk, particularly in optical wavelengths. This effect arises because the disk's central region is optically thicker than the limb, causing most of the light to be emitted from the central regions, making them appear brighter. In contrast, the limb region consists of upper and thinner layers, emitting less light than the central region, making the limb appear darker. Consequently, dark features (such as spots) and bright features (such as faculae, plages, and the network) imprint distinctive signatures in the time series of stellar brightness variations. \citet{Sasha_Eliana} and \citet{ElianaEA2020a} found that by characterizing the particular shape generated by faculae (M-like shape) or spot (V-like shape) transits recorded in the total solar irradiance (TSI) and comparing simultaneously with the Michelson Doppler Imager (MDI) observations, it is possible to infer whether faculae or spot regions dominated the stellar surface of solar analogs \cite[see also][]{ElianaEA2020b}. This method requires the analysis of the gradient of the power spectral (GPS) from photometric time series, which are available for thousands of stars. 

Overall, the current Sun is a faculae-dominated star. Thus, to directly compare solar-type stars to the Sun, between spot- and faculae-domination should be distinguished \citep[e.g.,][]{Sasha_Eliana, ElianaEA2020a}. Consequently, the Sun should only be compared to Sun-like stars of comparable masses, radii, effective temperatures, and rotation periods showing faculae-dominated characteristics.

Solar brightness fluctuates over periods ranging from minutes to decades. Notably, these variations in observed brightness can directly be linked to the Sun's rotation period. Understanding this relationship will help to make inferences about other stars. Despite successful stellar surveys by photometric missions like Kepler, the CHaracterising ExOPlanets Satellite \citep[CHEOPS, e.g.,][]{BenzEA2021}, and TESS, there remains a lack of photometric data on rotation periods for stellar solar analogs. The main challenges in accurately determining rotation periods for the Sun and its analogs include non-periodic light-curve profiles, low modulation amplitudes caused by the random appearance of magnetic features, and the brief lifespan of these features compared to the rotation period \citep{2017NatAs...1..612S,Sasha_Eliana}. 

Analyzed solar light curves utilizing regular periodograms as Lomb-Scargle, autocorrelation functions, or power spectra do not show a realistic rotation period value of (or close) to the expected 26-day period \cite[see][]{Aigrain,ElianaEA2020a}. This suggests that for stars with similar brightness variations to the Sun, similar issues in detecting their rotation periods might occur, and therefore implies that only a small fraction of solar-type systems have been thoroughly analyzed \cite[see,][]{2010ApJ...713L.155B,2020Sci...368..518R}.

Although there have been efforts to construct extensive stellar rotation rate catalogs, reliable information on rotation periods is still lacking (e.g., unknown for about 87\% of the 530,506 stars observed by the Kepler telescope \footnote{\href{https://science.nasa.gov/resource/nasas-kepler-mission-by-the-numbers/}{https://science.nasa.gov/resource/nasas-kepler-mission-by-the-numbers/}}). For solar twins, this difficulty is very challenging. Even after successfully deriving extensive stellar survey catalogs \citep[e.g.,][who published rotation periods of about 34,030 and 67,163 Kepler stars, respectively]{McQuillanEA2014,2023A&A...678A..24R}, there is a lack of solar twin systems adequately studied and characterized \citep[see,][]{ElianaEA2020b}. Thus, a detectability bias toward active stars -- with clearly sinusoidal and stable magnetic feature modulation --  prevents us from finding analog systems with similar magnetic and activity environments where similar planetary habitability conditions could exist.

The newly developed method discussed in \citet{ElianaThesis} allows to infer and quantify the degree of spot- or faculae-dominance on the stellar surface based on observed light curves (LCs) and revealed that Sun-like stars are distributed between three regimes: spot- and faculae-dominated stars or those transitioning between the two branches. The analysis of several Kepler Sun-like stars with the GPS method showed that their photospheres display a smooth transition between being dominated by spots or faculae, with the Sun sitting roughly in the middle \cite[][]{ElianaEA2020a}. This transition is quantified by the factor $\alpha$ corresponding to the instantaneous area ratio between dark or bright features ($S_{f}/S_{s}$) detected on the stellar surface. Note that in the case of the Sun, the $\alpha$ factor has a value of $\alpha_{\odot}$ = 0.158, which corresponds to a $\rm S_{fac}/S_{spot}$ ratio of $\sim 3$. This directly indicates that the current Sun, on a rotation time scale, is transitioning to a regime dominated by faculae.  

Surface manifestations of magnetic flux not only rely on parameters such as effective temperature and rotation period but also on the strength of the magnetic field, which, in turn, depends on the underlying dynamo. Small differences in the generation of internal magnetic fields will influence the surface features and their development into higher atmospheric layers, such as the chromosphere and the corona. Furthermore, these differences might impact the mechanism responsible for driving the stellar winds and the frequency of transient events like flares and coronal mass ejections (CMEs). These factors are crucial for understanding the evolution of stellar rotation and magnetic activity and critical for characterizing the environments in which exoplanets are embedded and for estimating the survival as well as conditions of their possible exoatmospheres.

\section{Magnetic energy stored around starspots}
\begin{figure}[!t]
\centering
\includegraphics[width=\columnwidth]{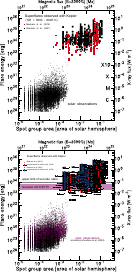}
\caption{Scatter plot of the flare energy as a function of spot group area
for solar flares (black dots) and superflares of G-type stars from the Kepler sample by \citet{OkamotoEA2021}. Left panel: Original Sun-like sample from \citet{OkamotoEA2021}. Right panel: Updated solar observations based on the rescaling discussed in \citet{hudson2024} (purple dots) and updated stellar sample utilizing the updated spot group area function by \citet{herbst2021starspots} (blue dots and according errors).}
\label{fig:1}
\end{figure}

In theory, the magnetic energy accumulated near starspots can explain the observed flare intensities. To put stellar and solar observations into context, often a scatter plot of the flare energy as a function of the starspot area is discussed. The upper panel of Fig.~\ref{fig:1} shows the solar flare intensities observed by the Geostationary Operational Environmental Satellites (GOES, black dots) and the superflare regime based on Kepler 1-min cadence observations \citep[red squares, see][]{MaeharaEA2015} and the sample obtained by \citet{OkamotoEA2021} (black squares). To generate the observed superflares, the existence of large starspots is an almost mandatory condition \citep[see discussion in][]{Notsu-etal-2019, OkamotoEA2021}, which is reflected in the much bigger spot group areas of the superflare regime. It further shows that the low-energy superflares of the short-cadence Kepler sample by \citet{MaeharaEA2015} are in the order of the maximum flare energies observed on the current Sun.

As discussed in \citet{2012Natur.485..478M}, \citet{ShibayamaEA2013}, and \citet{NotsuEA2013, Notsu-etal-2019}, the stellar area covered by a starspot can be described as a function of the stellar brightness variation and the ratio between the spot temperature and the stellar effective  temperature. However, \citet{herbst2021starspots} showed that the analytically derived stellar spot group areas are erroneous because the ratio between spot and effective temperature that previously has been proposed by \citet{BerdyuginaSV2005} needed to be updated. Further, \citet{hudson2024} showed that the solar flare energies measured by GOES-1 through GOES-15 (1975 – 2016) need to be scaled up by a factor of 1/0.7. An updated version of the plot incorporating the aforementioned corrections is shown in the lower panel of Fig.~\ref{fig:1}. These updates further lend additional support to the idea that superflares might happen on the Sun, as these corrections shift the solar observations well within the flare energy range of the 1-min cadence Kepler observations.

In fact, with the help of cosmogenic radionuclides like $^{10}$Be, $^{14}$C, and $^{36}$Cl \citep[e.g., ][]{Beer2012}, first hints for the occurrence of solar superflares within the past 25,000 years have been revealed: With extreme spike-like increases in all three archives - as of now - five extreme solar events have been identified: AD993, AD774/775, 660 BC, 5259 BC, and 7176 BC \citep[see, e.g.,][respectively]{2013NatCo...4.1748M, mekhaldi2015multi, O'hare-etal-2019,Brehm-etal-2021}. Note that three more events around 5410 BC, 1052 AD, and 1279 AD \citep[see][respectively]{10.1029/2021GL093419,2021NatGe..14...10B} have been found in the radionuclide records and that more may be detected in the future \citep[e.g.,][]{2025arXiv250205903H}. However, since these events, so far, have not been confirmed in all three records, they are not considered in this study. Nevertheless, one of the strongest (and most famous) ones is the AD774/775 event \citep[e.g.,][]{2023A&A...671A..66P, PapaioannouEA2024}. Assuming that the AD774/775 event was a single extreme solar event and based on the GOES re-calibration by \citet{hudson2024}, an associated X-ray flare intensity of 2$\times10^{33}$ - 6$\times10^{33}$ erg \citep[][]{2022LRSP...19....2C,2023A&A...671A..66P} has been derived (see purple horizontal band in the lower panel of Fig.~\ref{fig:1}), bringing the upper solar flare energies much closer to the upper limit of Sun-like stars based on the Kepler sample \citep[here from][]{OkamotoEA2021} and further supporting the consistency between solar and Sun-like star observations. 

In the following, the stellar sample by \citet{OkamotoEA2021} is reanalyzed with focus on spot- or faculae-domination of the Sun-like stars.

\section{Stellar sample used in this study}
In this study, as a first step, we utilized the Sun-like sample by \citet{OkamotoEA2021} to check the reported Sun-like stars for faculae domination.

Utilizing the GAIA/DR3 temperatures of the initial 265 solar-type main sequence stars, we found 119 to be within the effective temperature regime of 5600 K $< T_{\mathrm{eff}} <$ 6000 K. These 119 stars  form the background of our study.

\section{Facular versus spot dominance}
In this study we analyzed 119 stars in the field of view (FOV) observed during a 90-day quarter by the Kepler telescope and considered light curves acquired in the long-cadence mode (i.e., with a cadence of 29.42~min). We found that 48 stars showed a definitive signal of magnetic features transiting the stellar disk.

\begin{figure}[!t]
\centering
\includegraphics[width=\columnwidth]{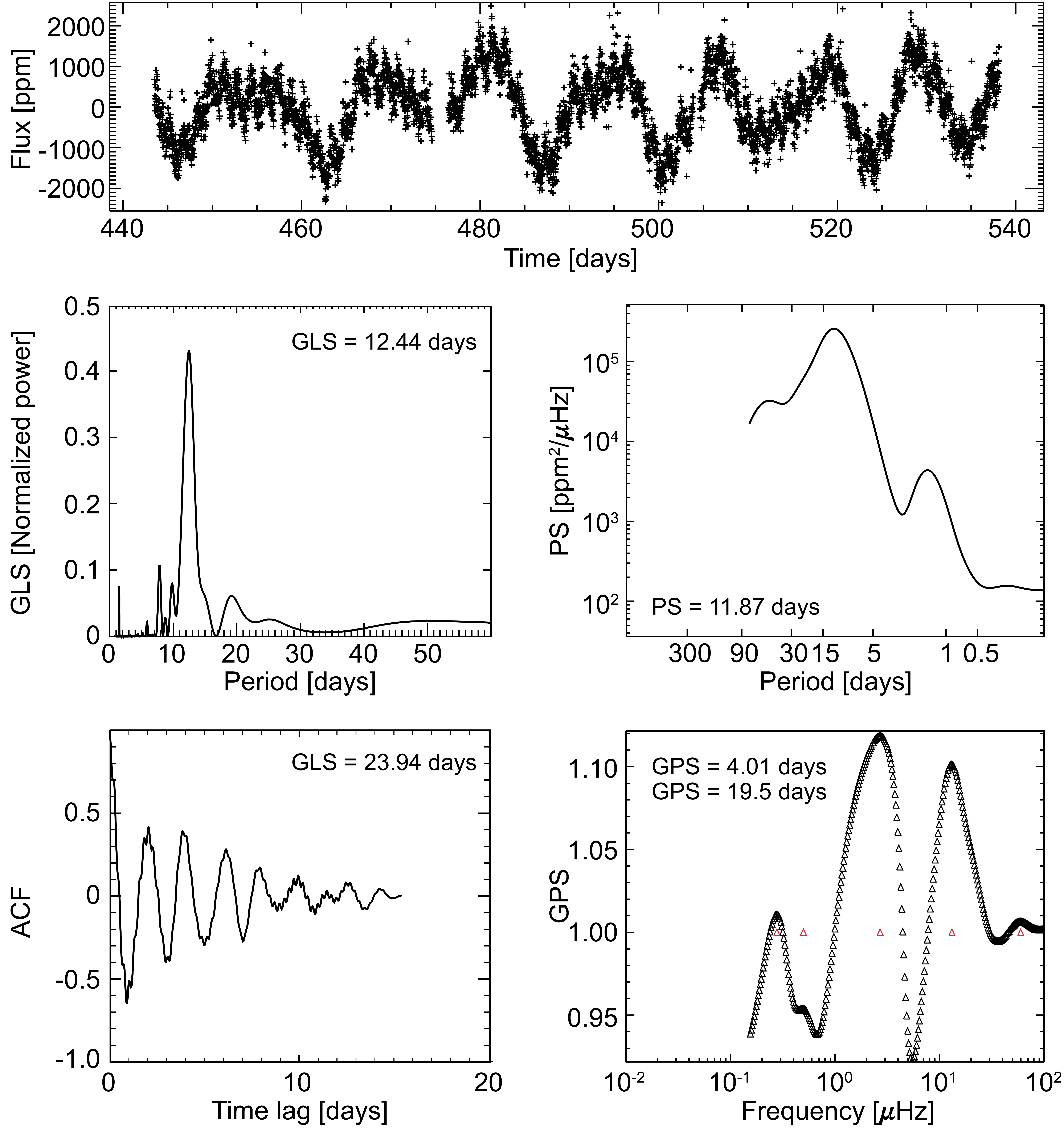}
\caption{Analysis of the rotation period based on the Kepler lightcurve of KIC\,11599385 (a). Generalized Lomb-Scargle GLS periodogram (b), autocorrelation function (ACF, c), power spectrum (d), and gradient of the power spectrum (GPS, d). Each panel displays the most prominent periods detected with each method.}
\label{fig:2}
\end{figure}

However, note that this high quality of the light curves was required to improve the ability to distinguish between surface features -- spot or faculae -- when applying the GPS method. A highly periodic stellar signal improves the definitive detection of a certain feature, an enhancement normally observed in stars that are more active than the Sun. This behavior is evident in the entire sample, where most of the analyzed stars show predominant periodic activity (with e.g., high bright contrast, longer than a single rotation lifetime evolution, and non-stochastic emergence of magnetic features). This results in a periodic light curve with a detectable M- or V-shape, which is characteristic for faculae or spots, respectively (see Fig.\ref{fig:2}-a for a periodic facular transit LC profile).

Such ideal periodic conditions have, however, only been observed on the Sun during isolated occasions (see \citealt{2016JSWSC...6A..30K}, and Figs. 1 and 2 in \citealt{ElianaEA2020a}). It is more common to observe a superposition of faculae and spots, where both phenomena coexist. Thereby, a solar LC is more disorganized in comparison with regular LCs of more active stars. Nevertheless, within a rotational time scale, on rare occasions, it is possible to observe a clear transit of isolated, stable (long-lasting), and periodic dark or bright features dominating the solar LC. A similar behavior is apparent in the LC of  KIC\,11599385 shown in Fig.~\ref{fig:2}(a). Since a higher contribution of bright features is observed in solar TSI, stronger faculae and network area components are normally detected. This phenomenology is the result of less stable and thinner emerging magnetic flux tubes, making the presence of bright features more evident \citep[][]{2000JApA...21..275F}. 

Within the time frame of a solar magnetic cycle, faculae are dominating. This explains why a brighter Sun is observed even during solar maximum conditions. Note that the brightness of the Sun in general varies only slightly with solar activity. Reconstructions of cosmogenic radionuclide production rates further indicate that the Sun is somewhat brighter today than within the past 8,000 years \citep[e.g.,][]{2004Natur.431.1084S, https://doi.org/10.1029/2009GL040142}, reflecting that the Sun likely had less facular contribution over the last few millennia. Thus, understanding the transition of stellar activity investigated in \citet{ElianaEA2020b} is crucial for forming a clear picture of the past, recent, and future changes of our Sun but also further to gather insight into the emergence and evolution of life in the solar system and beyond. 

This solar perspective gives a basic approach to the difficulties of finding solar twins in Sun-like stars and highlights the need for a careful comparison of the Sun to more periodic, more active, and then, easier to detect Sun-like stars. Further studies of low-periodicity and faculae-dominated stars are mandatory, and the scarcity of such stars in current observations may be due to observational challenges like low signal-to-noise ratios, long rotation periods, and the superposition of observable features.

In order to recover a characteristic signature from the surface features, we calculate the GPS as the ratio between the power spectral density $P(\nu)$ (see Fig.\ref{fig:2}(d)) at two adjacent frequency grid points ($GPS \equiv P(\nu_{k+1})/P (\nu_k)$) as follows:

\begin{equation}\label{eq:1}
GPS = 1+\frac{d \ln P(\nu_k)}{d \ln \nu} \cdot \frac{{(\Delta \nu)}_k}{\nu_k},
\end{equation}

where $\Delta \nu$ represents the spacing of the frequency grid, calculating the power spectra gradient on a equidistant logarithmic scale grid with $\Delta \nu / \nu$ assumed to be constant. After the GPS profile is derived (see, Fig.\ref{fig:2}(e)), we locate the highest amplitude peak that corresponds to the inflection points (IP$_{i}$) for the power spectrum profile. Therefore, the IP$_{i}$ values represent a gradient of the power spectrum plotted on a log-log scale. As discussed in \citet{1998BAMS...79...61T},\citet{Sasha_Eliana}, and \citet{ElianaThesis}, the latter depends on the chosen frequency grid. The location of the inflection points is proportional to the stellar-rotation period via the calibration factor defined as $\alpha_{i}=IP_{i}/P_{\rm rot}$, where IP$_{i}$ is the period corresponding to the ($i$) inflection point. In this case we only use the high-frequency, high-intensity inflection point and its corresponding $\alpha$ proportionality factor. Other IP peaks with lower frequency and intensity values are disregarded. Further, the uncertainty of the $\alpha$~value is computed as the standard deviation between the positions of the inflection points. 

Thus, from the light curve profile, the relative presence of faculae (M-like shapes) or spots (V-like shapes) results in a characteristic IP location in the GPS that allows the estimation of a proportionality factor $\alpha$ that describes quantitatively the presence of specific magnetic features on the stellar surface. We estimated the relative dependence of the facular-to-spot area ratio at the time of maximum area, $\rm S_{fac}/S_{spot}$, from the observed inflection point position and corresponding $\alpha$ values following the models discussed in \citet{Sasha_Eliana} and observed stars in \cite{ElianaEA2020b}.

\begin{figure}[!t]
\centering
\includegraphics[width=\columnwidth]{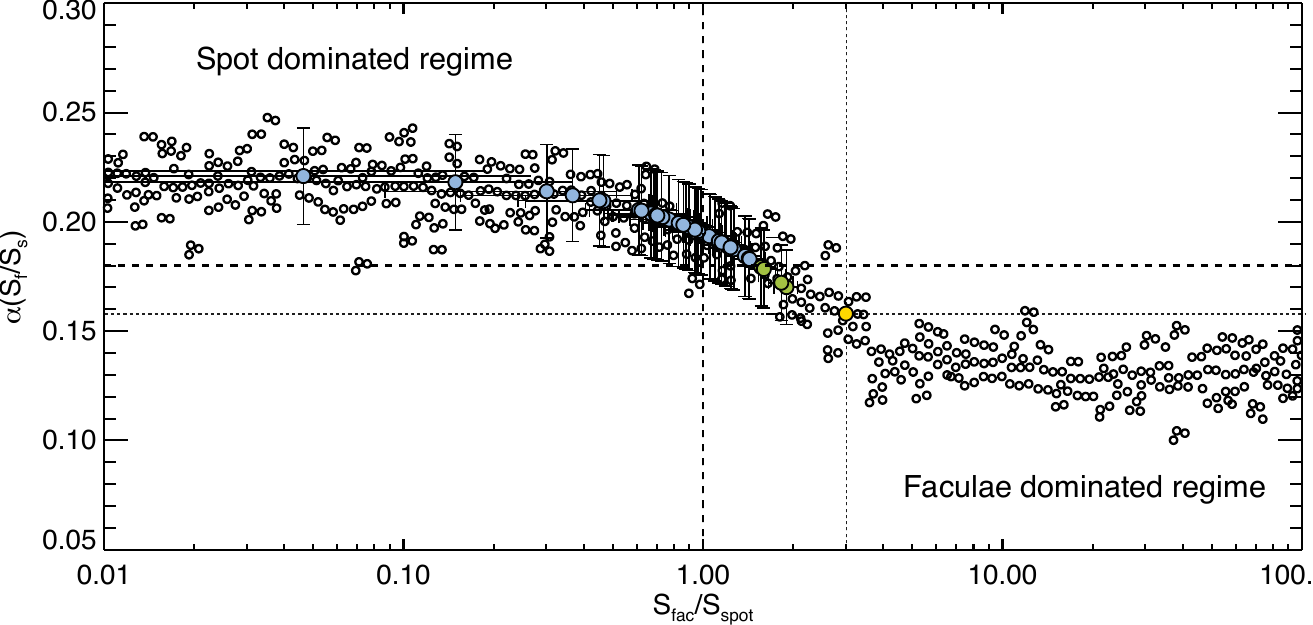}
\caption{Diagram of the $\alpha$-Factor (parameter proportional to the instantaneous surface ratio between bright and dark features, $\rm S_{fac}/S_{spot}$) for the 48 Kepler stars compared with values obtained for 400 modeled LC's with different $\rm S_{fac}/S_{spot}$ \citep[adapted from][]{Sasha_Eliana,ElianaEA2020b}. The Sun (yellow circle) is located in the transition region between the branches of spot-dominated surfaces (left) and faculae-dominated light curves (right). Blue circles represent the 48 stars analyzed with GPS in this work, four of which are highlighted in green (i.e., KIC\,3853938, KIC\,8424356, KIC\,11599385, and KIC\,12266582) since they are in the transition between being spot- and faculae-dominated stars (i.e., GPS $\alpha$<0.18\, as indicated by the dashed black line). The black dotted line corresponds to solar values.}
\label{fig:3}
\end{figure}

Figure~\ref{fig:3} shows the $\alpha$-factor for the 48 Kepler stars of our sample (blue/green dots) in comparison with values obtained for 400 modeled LCs with different $\rm S_{fac}/S_{spot}$ ratios \citep[adapted from][]{Sasha_Eliana,ElianaEA2020b}. As can be seen, the Sun ($\alpha_{\odot}$-Factor\,=\,0.158 and $\rm S_{fac}/S_{spot}\sim3$, yellow dot) is located in the transition region between the branches of spot-dominated surfaces (left) and faculae-dominated light curves (right). The latter indicates that the Sun is transitioning to a regimen dominated by faculae. As can be seen, only four of the 48 stars analyzed with GPS (highlighted in green, i.e., KIC\,3853938, KIC\,8424356, KIC\,11599385, and KIC\,12266582) are in the transition between being spot- and faculae-dominated stars (i.e., GPS $\alpha$<0.18\, as indicated by the dashed black line). Further, three of these five - within the estimated errors - show $\alpha$-values in agreement with the Sun (i.e., KIC\,8424356, KIC\,11599385, and KIC\,12266582).  However, within a rotational time scale, all the stars in the sample show a faster, periodic, and sinusoidal modulation in the LC. This implies that the selection of stars is far from the typical solar brightness behavior. The derived parameters of the 48 stars in our sample are listed in Tab.~\ref{tab:1}.

Note that the predominantly high variability of the Sun-like stars in our sample does not necessarily imply that the Sun is an outlier. A possible unbiased interpretation could be that by comparing solar variability of the sample of stars, the solar levels of photometric variability are typical for stars having near-solar fundamental parameters but unknown rotation periods.

\section{Conclusions and future prospects}
\begin{figure}[]
\centering
\includegraphics[width=\columnwidth]{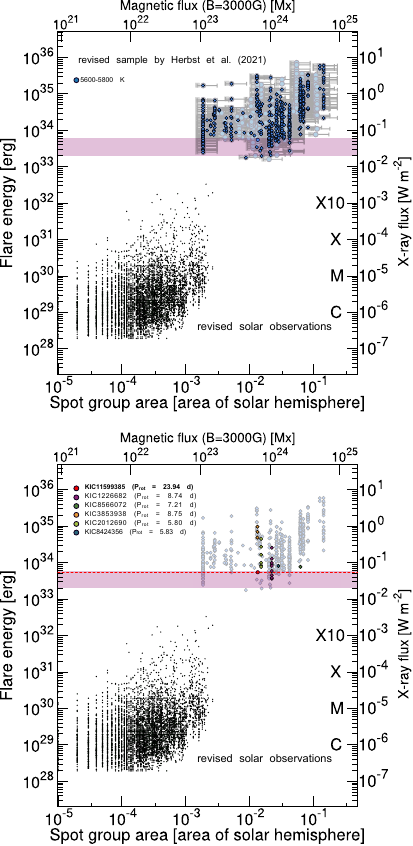}
\caption{Updated solar and stellar samples as shown in the right panel of Fig.~\ref{fig:1}. Upper panel: Highlighted in blue is the flaring of the 48 stars used in our analysis. Lower panel: Highlighted is the flaring activity of the four Sun-like stars transitioning between being spot- and faculae-dominated stars. The flare energy of the only star in the sample that further shows a comparable rotation period (i.e., KIC\,11599385) is highlighted by the red dashed line falling right within the reconstructions of the AD774/775 event.}
\label{fig:4}
\end{figure}
Out of the 48 stars, only four stars (i.e., KIC\,3853938, KIC\,8424356, KIC\,11599385, and KIC\,12266582) analyzed by the GPS were found to be in the transition between being spot- and faculae-dominated stars (i.e., GPS $\alpha$<0.18) and with that comparable to the  current Sun ($\alpha\,\sim$\,0.158), out of which only three roughly match the solar effective temperature, radius, and mass (i.e., KIC\,3853938, KIC\,8424356, and KIC\,11599385). 

However, only one star of the entire sample, namely KIC\,11599385, with a rotation period of $P_{rot}$\,=\,23.94 days, further matches the solar rotation period. Thus, in the given sample, KIC\,11599385 is the only true current Sun analog. During its observation, Kepler detected one flare that had a bolometric energy of 5.5$\times 10^{33}$ erg. As shown in Fig.~\ref{fig:4}, this flare energy falls right within the rescaled flare intensities of the AD774/775 event detected in the cosmogenic radionuclide records. 

Of course, 48 stars are not enough to fairly compare the Sun with its analogs, and follow-up studies are needed. Most recently, with the help of the Kepler archive\footnote{\href{https://exoplanetarchive.ipac.caltech.edu/docs/KeplerMission.html}{https://exoplanetarchive.ipac.caltech.edu/docs/KeplerMission.html}} and the Gaia DR3 release\footnote{\href{https://www.cosmos.esa.int/web/gaia/data-release-3}{https://www.cosmos.esa.int/web/gaia/data-release-3}} \citet{2024Sci...386.1301V} derived a new sample of main-sequence stars with effective temperatures of 5000 K $< T_{\mathrm{eff}} <$ 6500 K and absolute magnitudes between 4 mag $< M_G <$ 6 mag. They identified 2889 superflares on 2527 Sun-like stars. This new statistical analysis suggests that superflares are much more frequent on Sun-like stars than previously thought: for instance, superflares with energies $> 10^{34}$ erg on average likely occur roughly every 1000 years. The study further suggests that the resulting flare-frequency distribution fits an extrapolation of the solar distribution indicative of the same physical mechanisms. The latter most recently has been investigated by \citet{2025A&A...695A..21Y}, who found that fast rotators have the same dynamo process as the Sun and that a solar-type dynamo can be assumed at all evolutionary stages of a star.

Therefore, in a next step, we will utilize the newly derived Kepler sample by \citet{2024Sci...386.1301V} and add available TESS data to redo the analysis presented above. 

However, even though we searched for similar T$_{eff}$, P$_{rot}$, stellar ages, and faculae-dominated stars (or those in transition to faculae domination), a difference in stellar metallicity will change the picture. As discussed in \citet{WitzkeEA18}, even a slight variation in metallicity or effective temperature, for example, within the observational error range, substantially impacts the photometric brightness change in relation to the Sun. Thus, it is essential to precisely ascertain the fundamental stellar properties to fully understand variations in stellar brightness.

Undoubtedly, the Kepler mission brought long-term time series for thousands of Sun-like stars. However, the specific location of its relatively faint field of view (FoV) restricted a proper ground-based spectroscopic follow-up, limiting comprehensive spectral information for many targets. The lack of complementary information for the Kepler field straightened the robust characterization of the solar analogs. For the extensive TESS survey, spectral follow-up has been possible. However, the 27-day averaged short light curves represent an issue in obtaining the rotational modulation for closer solar analogs with a slow rotation period compared to the Sun.

Within the following years, it is expected that the support activities for the PLATO mission \citep{RauerEA2014,RauerEA2024} can provide long-term follow-up observations of stellar activity, particularly of G-stars, and will improve, for example, the sampling of targets with simultaneous or contemporaneous spectral information for the upcoming PLATO high-cadence and long-term photometric time series. The measuring of stellar surface rotation, photometric activity, and long-term modulations is an integral part of the PLATO pipeline \citep{2024A&A...689A.229B}. In addition, the mission shall be able to deliver more accurate and precise stellar characterization for a larger sample of targets than Kepler~\citep{2024A&A...683A..78G}, hence improving our insight into the science topics discussed in this paper.

\begin{acknowledgements}
KH acknowledges the support of the DFG priority program SPP 1992 “Exploring the Diversity of Extrasolar Planets (HE 8392/1-1)”. EMAG was partially supported by HST GO-15299 and GO-15512 grants and acknowledges support from the German \textit{Leibniz-Gemeinschaft} under project number P67/2018. AP acknowledges the support from NASA/LWS project NNH19ZDA001N-LWS. KH and AP acknowledge the International Space Science Institute and the supported International Team 464 (ETERNAL). This work presents results from the European Space Agency (ESA) space mission PLATO. The PLATO payload, the PLATO Ground Segment and PLATO data processing are joint developments of ESA and the PLATO Mission Consortium (PMC). Funding for the PMC is provided at national levels, in particular by countries participating in the PLATO Multilateral Agreement (Austria, Belgium, Czech Republic, Denmark, France, Germany, Italy, Netherlands, Portugal, Spain, Sweden, Switzerland, Norway, and United Kingdom) and institutions from Brazil. Members of the PLATO Consortium can be found at \href{https://platomission.com/}{https://platomission.com/}. The ESA PLATO mission website is \href{https://www.cosmos.esa.int/plato}{https://www.cosmos.esa.int/plato}. We thank the teams working for PLATO for all their work. This project further has received funding from the Research Council of Norway through the Centres of Excellence funding scheme, project number 332523 (PHAB) and was supported by the European Union (ERC, PastSolarStorms, grant agreement no. 101142677). Views and opinions expressed are however those of the author(s) only and do not necessarily reflect those of the European Union or the European Research Council. Neither the European Union nor the granting authority can be held responsible for them.

\end{acknowledgements}


\bibliographystyle{aa}
\bibliography{references}

\begin{appendix}
\onecolumn
\begin{landscape}
\section{Stellar sample}\label{appendix:A}
\centering
\begin{longtable}{lcccccccccccccc}
\caption{Characteristics of the stellar sample and corresponding parameters derived in this study.}\label{tab:1}\\
\label{lsltapp} 

KIC ID & $T_{\mathrm{eff}}$ & $\log g$ & [Fe/H] & R & Quarter & INF & Time 
[d] & Period[d] & PS [d] & ACF [d] & LS [d] & VAR [ppms] & $\alpha_\star$ & F/S \\
\hline
\endfirsthead
\caption{continued.}\\
\hline\hline
KIC ID & $T_{\mathrm{eff}}$ & $\log g$ & [Fe/H] & R & Quarter & INF & Time 
[d] & Period[d] & PS [d] & ACF [d] & LS [d] & VAR [ppms] & $\alpha_\star$ & F/S \\
\hline
\endhead
\hline
\endfoot
\kh{3853938} & \kh{5790} & \kh{4.396} & \kh{-0.18} & \kh{0.998} &  \kh{7} & \kh{7.3739386} & \kh{1.5695919} & \kh{7.8479594} & \kh{8.6886948} & \kh{8.7540477} & \kh{9.101} & \kh{8862.8539} & \kh{0.17929899} & \kh{1}\footnote{Stars in the transition between being spot- and faculae-dominated stars are highlighted in blue.} \\
4142137 & 5761 & 4.561 & -0.26 & 0.828 & 7 & 5.214162 & 2.2197381 & 11.098691 & 11.026396 & 11.555357 & 11.8977 & 17076.826 & 0.19209601 & 0 \\
5263650 & 5810 & 4.447 & -0.84 & 0.842 & 6 & 3.6079651 & 3.2079229 & 16.039615 & 15.593679 & 15.771153 & 16.019 & 8621.7163 & 0.20340446 & 0 \\
5374789 & 5671 & 4.431 & -0.42 & 0.896 & 9 & 5.6860839 & 2.0355088 & 10.177544 & 9.2720573 & 9.5130938 & 9.72872 & 12212.429 & 0.21396918 & 0 \\
5459300 & 5745 & 4.426 & -0.16 & 0.962 & 4 & 3.8483217 & 3.0075641 & 15.03782 & 13.119088 & 15.874217 & 7.79166 & 9491.1095 & 0.1894622 & 0 \\
5563561 & 5927 & 4.525 & -0.58 & 0.831 & 6 & 12.672993 & 0.91328654 & 4.5664327 & 4.4394762 & 4.5420923 & 4.37592 & 15320.699 & 0.20107177 & 0 \\
5695372 & 5675 & 4.569 & -1.28 & 0.698 & 7 & 10.890025 & 1.0628143 & 5.3140713 & 5.1663289 & 5.3285161 & 5.15545 & 12851.411 & 0.19945783 & 0 \\
5953631 & 5644 & 4.663 & -0.88 & 0.67  & 5 & 17.913551 & 0.64610721 & 3.230536 & 3.1407203 & 3.2917313 & 3.15878 & 16031.701 & 0.19628188 & 0 \\
5991070 & 5778 & 4.087 & -0.20 & 1.452 & 9 & 9.3533256 & 1.2374288 & 6.1871438 & 6.1468423 & 6.3394607 & 6.27402 & 16773.677 & 0.19519464 & 0 \\
6059055 & 5772 & 4.528 &  0    & 0.901 & 6 & 4.1986787 & 2.7565991 & 13.782996 & 13.399799 & 13.626391 & 13.8011 & 14681.211 & 0.20229855 & 0 \\
6352768 & 5674 & 4.582 & -0.38 & 0.782 &  8 & 9.7722136 & 1.1843861 & 5.9219306 & 4.9472935 & 5.0486163 & 4.73738 & 14147.002 & 0.23459618 & 0 \\
6431380 & 5986 & 4.481 & -0.14 & 0.954 & 7 & 4.2906178 & 2.6975309 & 13.487654 & 13.993059 & 14.221846 & 14.8683 & 8778.2892 & 0.18967516 & 0 \\
6610891 & 5815 & 4.625 & -0.8 & 0.718  & 6 & 15.738056 & 0.73541957 & 3.6770979 & 3.8984199 & 3.9112968 & 1.93335 & 9454.9482 & 0.18802449 & 0 \\
6613812 & 5757 & 4.544 & -0.08 & 0.872 & 9 & 4.1066996 & 2.8183396 & 14.091698 & 14.306466 & 14.586726 & 14.9672 & 9812.7275 & 0.19321263 & 0 \\
7287601 & 5787 & 4.396 & -0.08 & 1.018  & 5 & 2.607081 & 4.4394762 & 22.197381 & 21.580247 & 21.780906 & 22.4925 & 6849.8984 & 0.20382422 & 0 \\
7293816 & 5776 & 4.456 & -0.14 & 0.942  & 6 & 5.6860839 & 2.0355088 & 10.177544 & 9.2720573 & 9.2124895 & 9.33343 & 13162.693 & 0.22095101 & 0 \\
7354508 & 5638 & 4.391 & -0.42 & 0.928  & 7 & 8.2174194 & 1.4084804 & 7.042402 & 6.5563346 & 6.7241385 & 6.86077 & 13326.849 & 0.2094663 & 0 \\
7869831 & 5609 & 4.513 & -0.04 & 0.881  & 6 & 3.1003593 & 3.7331397 & 18.665698 & 17.757905 & 17.790078 & 17.9415 & 16101.581 & 0.20984392 & 0 \\
8042739 & 5960 & 4.583 & -0.58 & 0.793  & 6 & 10.890025 & 1.0628143 & 5.3140713 & 5.0556251 & 5.1729725 & 5.03971 & 16089.189 & 0.20545523 & 0 \\
8167703 & 5937 & 4.373 & -0.58 & 0.973  & 9 & 5.1024332 & 2.2683441 & 11.34172 & 10.11125 & 10.398907 & 10.4589 & 11458.559 & 0.21813293 & 0 \\
8285970 & 5761 & 4.561 & -0.26 & 0.828  & 9 & 4.9930986 & 2.3180143 & 11.590072 & 11.026396 & 11.413458 & 11.5795 & 15631.577 & 0.20309482 & 0 \\
\kh{8424356} & \kh{5747} & \kh{4.344} & \kh{-0.24} & \kh{1.036}  & \kh{9} & \kh{11.123041} & \kh{1.0405494} & \kh{5.202747} & \kh{5.6366792} & \kh{5.8322718} & \kh{5.85824} & \kh{18346.996} & \kh{0.17841237} & \kh{1} \\
9049540 & 5880 & 4.518 & -0.06 & 0.917  & 9 & 5.8077502 & 1.9928671 & 9.9643353 & 9.0778174 & 8.8819493 & 9.25586 & 3659.4336 & 0.22437271 & 0 \\
9050543 & 5720 & 4.454 & -0.72 & 0.837  & 10 & 12.129763 & 0.95418802 & 4.7709401 & 4.6382979 & 4.7203317 & 4.80714 & 14645.892 & 0.20214427 & 0 \\
9138848 & 5821 & 4.348 & -0.2 & 1.055   & 6 & 4.6766628 & 2.4748575 & 12.374288 & 12.562881 & 12.616959 & 12.8152 & 13339.037 & 0.19615325 & 0 \\
9141747 & 5807 & 4.524 & 0.07 & 0.922   & 6 & 4.1087097 & 2.8169608 & 14.084804 & 13.693217 & 13.884996 & 14.4551 & 10951.5 & 0.20287805 & 0 \\
9410906 & 5696 & 4.329 & 0.14 & 1.126   & 7 & 5.9349234 & 1.950164 & 9.7508201 & 9.8994301 & 10.403351 & 10.3709 & 9790.7056 & 0.18745537 & 0 \\
9509071 & 5884 & 4.319 & -0.1 & 1.126 & 9 & 2.3895342 & 4.8436529 & 24.218264 & 23.040426 & 22.831556 & 11.3125 & 4264.5522 & 0.2121473 & 0 \\
9779373 & 5911 & 4.404 & -0.26 & 0.997  & 10 & 3.7658601 & 3.0734211 & 15.367106 & 15.267008 & 15.146839 & 15.5465 & 7152.1474 & 0.20290842 & 0 \\
10064358 & 5755 & 4.388 & -0.1 & 1.016   & 9 & 2.6641686 & 4.3443474 & 21.721737 & 22.535686 & 22.593144 & 22.6109 & 3058.51 & 0.19228609 & 0 \\
10067449 & 5685 & 4.011 & 0.21 & 1.73   & 6 & 6.9099985 & 1.6749749 & 8.3748745 & 8.502514 & 8.4534756 & 8.88196 & 8044.4981 & 0.19814038 & 0 \\
10287183 & 5912 & 3.978 & -0.94 & 1.504  & 7 & 10.884697 & 1.0633345 & 5.3166724 & 4.4416493 & 4.5673418 & 4.57385 & 24214.904 & 0.23281255 & 0 \\
10351715 & 5659 & 4.567 & -0.48 & 0.775  & 6 & 3.9345139 & 2.9416783 & 14.708391 & 15.593679 & 15.518944 & 16.3104 & 7160.4699 & 0.18955403 & 0 \\
10529366 & 5741 & 4.556 & -0.24 & 0.833  & 3 & 5.5642429 & 2.0800807 & 10.400403 & 10.332658 & 10.753563 & 10.8705 & 13603.166 & 0.19343177 & 0 \\
10801794 & 5776 & 4.512 & 0.21 & 0.95  & 6 & 8.3932491 & 1.3789742 & 6.894871 & 5.5158968 & 5.6789511 & 5.6353 & 15501.203 & 0.24282199 & 0 \\
10864581 & 5851 & 4.448 & -0.36 & 0.923   & 5 & 3.6062 & 3.2094931 & 16.047466 & 16.292041 & 16.335713 & 16.5734 & 4170.0182 & 0.19647095 & 0 \\
 10936708 & 5822 & 4.542 & -0.46 & 0.819  & 6 & 9.5628152 & 1.2103208 & 6.051604 & 6.143835 & 6.1823701 & 6.59614 & 13180.95 & 0.19576971 & 0 \\
 11017553 & 5778 & 3.921 & -0.44 & 1.763   & 6 & 7.8690278 & 1.4708391 & 7.3541957 & 7.3062923 & 7.4439773 & 7.47552 & 11153.455 & 0.19758781 & 0 \\
 11141091 & 5806 & 4.528 & -0.04 & 0.899 & 6 & 3.0324406 & 3.8167521 & 19.08376 & 18.959453 & 19.430131 & 19.5014 & 11026.678 & 0.1964347 & 0 \\
 11403440 & 5648 & 4.498 & 0.16 & 0.935  & 8 & 5.6833021 & 2.0365052 & 10.182526 & 10.564082 & 11.129414 & 11.318 & 10397.586 & 0.18298404 & 0 \\
 11413690 & 5959 & 4.277 & -0.1 & 1.202 & 6 & 7.6966434 & 1.503782 & 7.5189102 & 7.6335041 & 7.5702439 & 8.54356 & 6849.8019 & 0.1986438 & 0 \\
 \kh{11599385} & \kh{5636} & \kh{4.266} & \kh{-0.2} & \kh{1.119}  & \kh{5} & \kh{2.8416511} & \kh{4.0730103} & \kh{20.365052} & \kh{11.772473} & \kh{23.944019} & \kh{12.4354} & \kh{2491.4712} & \kh{0.17010554} & \kh{1} \\
 11652015 & 5794 & 4.118 & 0.14 & 1.496  & 5 & 4.1966246 & 2.7579484 & 13.789742 & 9.6873057 & 10.001791 & 10.1556 & 12526.296 & 0.27574545 & 0 \\
 11658610 & 5967 & 4.419 & -0.08 & 1.025 & 10 & 3.3069 & 3.4999771 & 17.499885 & 14.306466 & 18.302305 & 14.5748 & 2698.2028 & 0.19123149 & 0 \\
 11666355 & 5850 & 4.556 & -0.3 & 0.838  & 4 & 4.0187015 & 2.8800532 & 14.400266 & 13.406358 & 14.040841 & 13.7885 & 9885.6227 & 0.2051197 & 0 \\
 \kh{12266582} & \kh{5981} & \kh{4.439} & \kh{-0.7} & \kh{0.896}  & \kh{5} & \kh{7.6966434} & \kh{1.503782} & \kh{7.5189102} & \kh{8.5066758} & \kh{8.7358214} & \kh{8.82693} & \kh{10503.931} & \kh{0.17213974} & \kh{1} \\
 12453480 & 5658 & 4.536 & -0.72 & 0.765 & 7 & 11.128486 & 1.0400403 & 5.2002017 & 5.2794567 & 5.4581111 & 5.44108 & 10142.697 & 0.1905495 & 0 \\
 12456337 & 5763 & 4.133 & -0.16 & 1.37  & 8 & 4.4805797 & 2.5831644 & 12.915822 & 13.112669 & 13.717737 & 14.5167 & 4031.0796 & 0.18830835 & 0\\
\end{longtable}
\end{landscape}
\FloatBarrier 
\clearpage

\end{appendix}
\end{document}